\documentstyle[12pt,aaspp4]{article}

\begin{document}

\title{Spatial/Spectral Resolution of a Galactic Bulge K3 Giant
Stellar Atmosphere via Gravitational Microlensing}

\author{
Sandra Castro\altaffilmark{1},
Richard W.\ Pogge\altaffilmark{2},
R. Michael Rich\altaffilmark{3},\\
D.\ L.\ DePoy\altaffilmark{2},
Andrew Gould\altaffilmark{2,4}
}

\altaffiltext{1}{Department of Astronomy, Caltech, Pasadena, CA, 91125.
email: smc@astro.caltech.edu}
\altaffiltext{2}{Department of Astronomy, The Ohio State University, 
140 W. 18th Ave, Columbus, OH 43210-1173.  email: pogge, depoy, 
gould@astronomy.ohio-state.edu}
\altaffiltext{3}{Department of Physics \& Astronomy, University of California,
Los Angeles, CA, 90095.  email: rmr@astro.ucla.edu}
\altaffiltext{4}{Laboratoire de Physique Corpusculaire et Cosmologie,
College de France, 11 pl.\ Marcelin Berthelot, F-75231, Paris, France}

\begin{abstract}

We present two Keck HIRES spectra ($\lambda/\Delta \lambda=40,000$) of a
bulge K3 giant taken on successive nights during the second caustic
crossing of the binary microlensing event EROS BLG-2000-5.  This caustic
crossing served effectively to resolve the surface of the source star:
the spectrum from the second night is dominated by the limb, while the
spectrum from the first night is comprised of light from a broader range
of radii.  To demonstrate that the spectra are adequate to resolve the
differences between them, we analyze the H$\alpha$ line.  The equivalent
width is $\sim8\%$ smaller on the second night, and the signal-to-noise
ratio per resolution element (165 and 75 on the two nights respectively)
is sufficient to show that the difference is approximately constant over
the $\sim 2$\,\AA\ ($\sim15$ resolution element) extent of the line.
The sign of the difference is in the expected direction since the limb
is the coolest part of the star and therefore should have the weakest
H$\alpha$.  We invite atmosphere modelers to predict the difference
spectrum from the entire spectral range $5500<\lambda<7900\,$\AA\ so
that these predictions can be compared to our observations.

\end{abstract}

\keywords{gravitational lensing -- stars: atmospheres -- stars: imaging
-- stars: individual (EROS BLG-2000-5)}
 
\section{Introduction}

The spectrum of the Sun varies as a function of position on the Sun's
disk.  The surface of last scattering lies deeper in the atmosphere at
the center than it does at the limb.  The center therefore appears
hotter, and therefore brighter and bluer.  Spectral lines measured near
the center (compared to those at the limb) reflect the hotter conditions
at the surface of last scattering and also bear the imprint of a
different pattern of absorption features imposed upon the spectrum as
the light traverses a different path through the atmosphere.

These same effects are presumably present in all stars and are routinely
taken into account when modeling stellar atmospheres.  However, they
have never been directly observed in any star except the Sun and nearby
supergiants like Betelgeuse (e.g., Uitenbroek, Dupree \& Gilliland 1998),
because the angular sizes of other stars are generally too small to take
spectra at different spatial positions.  Limb darkening coefficients
have also been derived for a handful of eclipsing binary stars from a
combination of spectroscopy and light curves (e.g., GZ Cma; Popper et
al.\, 1985).  Hence, while spatially resolved models of stellar
atmospheres are well tested for solar type stars, they are poorly tested
for most other types of stars, in particular giants and metal-poor
stars.  For these, models and observations can be compared only for the
{\it integrated} spectra.

Microlensing can be used to measure limb darkening.  Microlensing
magnification patterns have caustics, (infinitesimally small) loci of
formally infinite magnification.  If the source passes over a caustic,
then at different times different parts of the source are highly
magnified, so from a time series of flux measurements one can
reconstruct the stellar brightness profile.

When Witt (1995) first proposed this idea, he specifically considered
microlensing by a point lens.  In fact, all four limb-darkening
measurements made to date (Albrow et al.\ 1999, 2000, 2001; Afonso et
al.\ 2000) used binary microlensing events.\footnote{Alcock et al.\
(1997) marginally detected limb darkening in a point-lens event, but could
not measure a limb-darkening coefficient.}  This is no accident.  Point
lenses have point-like caustics, so the cross section for a caustic
crossing is the angular diameter of the source.  By contrast, binary
lenses have one to three closed linear caustics with total cross
sections that tend to be more than two orders of magnitude larger than
the source.  Thus, even though recognizable binary microlensing events
are relatively rare, they comprise the great majority of caustic
crossing events.

Valls-Gabaud (1996, 1998) showed that stellar atmospheres could be
resolved spectrally by applying the same principle to a time series of
spectra rather than broad-band measurements.  Heyrovsk\'y, Sasselov, \&
Loeb (2000) made detailed predictions of how such spectra might vary
during a microlensing event.  Several other closely related applications
of time series of microlensing spectra were presented by Maoz \& Gould
(1994), Gould (1997), and Ignace \& Hendry (1999).

All of these theoretical investigations focused on point lenses.  Gaudi
\& Gould (1999) argued that to be practical, spatial/spectral resolution
would require binary-lens caustics.  The most important reason is that
binary caustic crossings can be accurately predicted in advance: binary
caustics are closed curves, so once the caustic region is entered, there
must inevitably be an exit crossing.  The approach to this crossing is a
square-root singularity which can be modeled without detailed knowledge
of the (generally complex) full binary-lens light curve.  Gaudi \& Gould
(1999) made a detailed study of the signal-to-noise ratio (S/N)
characteristics of such observations.  Their most important conclusion
in this regard was that observations during the second half of the
crossing (between the times when the center and the trailing limb exits
the caustic) carry substantially more spatial-resolution information
than those during the first half (see their Figure 4b).

\begin{figure}
\plotone{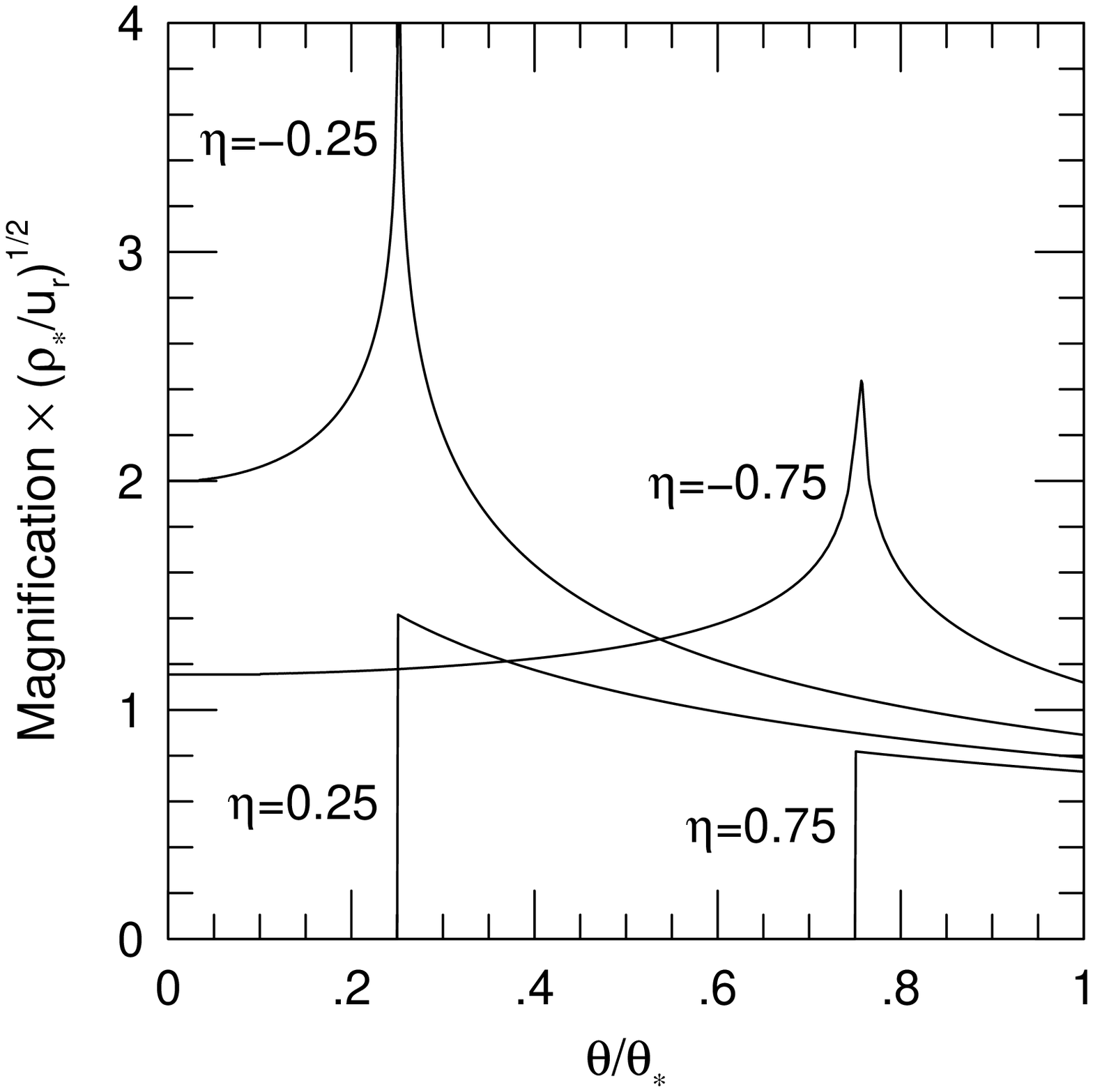}
\caption[junk]{\label{fig:one} Relative magnification averaged over
concentric stellar radii, $\theta$, at various phases, $\eta$, due to the
two extra images during a binary-lens caustic crossing.  Here $\eta$ is
the separation between the center of the source and the caustic in units of
the source radius, $\theta_\ast$.  During the second half of the crossing
($\eta>0$), there is a sharp cutoff in the magnification for
$\theta/\theta_\ast<\eta$.  The stellar profile can be deconvolved from a
series of spectra taken during this interval.  During the first half of the
crossing $(\eta<0)$, there are features in the magnification profile at
$\theta/\theta_\ast=-\eta$, but these are much less pronounced, so
deconvolution would be much more difficult.  See also Fig.\ 4b from Gaudi
\& Gould (1999).  The magnification profile is proportional to the
parameter combination $(\rho_\ast/u_r)^{-1/2}$.  Typically $10^{-2}\ga
\rho_\ast/u_r\ga 10^{-3}$, so the magnifications are scaled up from this
figure by a factor 10--30.}
\end{figure}

The fundamental reason for this can be understood from Figure
\ref{fig:one}.  When the source is inside a caustic, it has five
images, so the magnification can be written $A=A_2 + A_3$, where $A_2$
is the magnification of the two images that brighten divergently (and
then disappear) as the source approaches the caustic, and $A_3$ is the
magnification of the three other images.  To a good approximation, $A_3$
can be regarded as a constant during the crossing, and
\begin{equation}
A_2 =\biggl({\rho_\ast\over u_r}\biggr)^{-1/2} 
\biggl(-{\theta_\perp\over\theta_\ast}\biggr)^{-1/2} \Theta(-\theta_\perp).
\label{eqn:atwo}
\end{equation}
Here $\theta_\perp$ is the angular separation of a given position in the
source plane from the caustic, $\theta_\ast$ is the angular source size,
$\rho_\ast=\theta_\ast/\theta_{\rm E}$, $\theta_{\rm E}$ is the angular
Einstein radius, $u_r$ is a parameter that depends on the details of the
lens geometry, and $\Theta$ is a step function.  Typically $u_r\sim 1$,
$\rho_\ast\la 10^{-2}$, and $A_3$ is of order a few.  Hence, during a
caustic crossing, the magnification is dominated by $A_2$.  Figure
\ref{fig:one} shows the relative magnification due to $A_2$ of different
concentric rings on the stellar surface during different phases of the
crossing.  These phases are characterized by $\eta$, the angular distance
from the center of the source to the caustic in units of $\theta_\ast$.
Note that for the two curves with $\eta>0$ (second half of the crossing),
the $A_2$ images sample roughly uniformly the outer part of the star
($\theta/\theta_\ast>\eta$), but do not sample the inner part at all.  By
taking a series of such spectra, one can deconvolve the spatial profile.
By contrast, while the $\eta<0$ curves possess a sharp feature near
$\theta/\theta_\ast=-\eta$, its effect is muted by the high and roughly
uniform magnification from the rest of the star.  Here $\theta/\theta_\ast$
is the fractional distance of a given concentric ring from the center of
the star.  Note that this formalism cannot be applied to cusp-crossing
events or more generally to events where some parts of the source are
roughly equidistant from two caustics while other parts are crossing a
caustic.  These constitute a significant fraction of observed
caustic-crossing events (e.g., Alcock et al.\ 2000) and must be handled
differently (e.g., Albrow et al.\ 1999).

Alcock et al.\ (1997) acquired spectra of the high-magnification event
MACHO 95-BLG-30 and observed changes in H$\alpha$ and TiO near 6700\AA\
suggestive of center-to-limb variations in the spectral lines.  Lennon
et al.\ (1996) attempted to observe spectroscopic changes during the
second caustic crossing of the binary microlensing event MACHO 96-BLG-3
using the ESO NTT spectrograph, but failed to see any.  They concluded
that 8-10 m class telescopes would be required.

In this Letter we present high-resolution spectroscopy of a
caustic-crossing microlensing event of a K3 giant.  We demonstrate that
we have indeed resolved the stellar atmosphere by comparing the
H$\alpha$ absorption lines in spectra taken on successive nights during
the second half of the crossing.  We show that H$\alpha$ absorption is
significantly smaller in equivalent width on the second night when the
magnified light is dominated by the limb.  This is qualitatively
expected, since the limb is cooler than the center, and therefore
effectively corresponds to a later spectral type that has weaker
hydrogen lines.

The spectra contain a vast wealth of information.  To extract the
quantitative implications for the atmosphere requires a comparison of
the data with the predictions from detailed atmospheric models convolved
with the precise magnification profiles at the times of the observations
as determined from detailed modeling of the photometric light curves,
neither of which are available at present.  However, we invite modelers
to make their predictions so that they may be compared with the
observations in a future paper.

\section{Observations}

The EROS collaboration issued an alert on 2000 April 29 that EROS
BLG-2000-5 was a probable microlensing event with a clump giant source
toward the Galactic bulge
(www-dapnia.cea.fr/Spp/Experiences/EROS/alertes.html).  On 2000 June 8,
the MPS collaboration (bustard.phys.nd.edu/MPS/index.html) issued an
anomaly alert, noting rapid source brightening.  The PLANET
collaboration (thales.astro.rug.nl/$\sim$planet) then began intensive
observations within 4 hours of the onset of the first crossing,
measuring its duration to be $2\Delta t_1=1.0\,$days.

PLANET issued further anomaly alerts, identifying the source as a K3
giant and predicting the date of the second crossing.  By real-time
modeling of their observations, PLANET predicted that the source would
cross the second caustic at a much more acute angle than the first, and
therefore the second crossing would last several days.\footnote{For
sufficiently acute crossings, especially near a cusp, the approximation
of constant $u_r$ (embodied in eq.\ [1]) can break down.  This has no
effect on the arguments and results presented here, but it must be taken
into account in detailed modeling of the event.  For EROS BLG-2000-5 the
approximation is likely to remain valid.}

This extremely long crossing set the stage for Keck observations which
otherwise would probably not have been feasible.  Normally, the duration
of the second-half crossing (which Gaudi \& Gould 1999 showed contains
most of the information) is only a few hours.  The chance that it will
be visible from a given observatory site is therefore small, and the
chance that it will be visible over a wide range of $\eta$ (needed to
obtain {\it differential} measurements of the stellar atmosphere) is
smaller still.  However, a crossing that lasts several days can be seen
at substantially different $\eta$ simply by viewing it on different
nights.  In their last alert prior to the onset of the observations
described below, PLANET predicted that the center of the source would
cross the second caustic on HJD=2451731.6, and predicted that the
duration of the second-half crossing would be about 2 days.  They urged
immediate spectroscopic observations, which we then undertook.

\subsection{Spectroscopic Observations}

We obtained spectra of the bulge K3 giant star being lensed by EROS
BLG-2000-5 on UTC 2000 July 6 and 7 using the High Resolution
Spectrograph (HIRES; Vogt 1994) with the W. M. Keck 10-m telescope.  The
spectra cover the range $5500<\lambda<7900$\AA\ at resolution
$\lambda/\Delta\lambda\approx 40,000$.  On July 6 we took three
exposures of 1800\,s starting at 09:57:26 UTC, and on July 7 we took
four 1800\,s exposures starting at 09:35:48 UTC.  Total integration
times were 5400\,s and 7200\,s on the respective nights.  The
mid-exposure Heliocentric Julian dates of the combined spectra are
HJD=2451731.953 and HJD=2451732.950, respectively.  The spectra of July
6 were taken in 0\farcs8 seeing through light cirrus using the C1 slit
decker.  This gave slit dimensions of 0\farcs86$\times$7\farcs0,
providing a resolution of $\sim$45,000.  On the second night the seeing
was 1\farcs2--1\farcs4 through heavy variable cirrus clouds.  To improve
the S/N of the spectra we used the C5 slit decker which has slit
dimensions of 1\farcs15$\times$7\farcs0, giving a resolution of
$\sim$34,000.  The S/N ratios in the H$\alpha$ echelle order are
$\sim90$ and $\sim40$ per pixel (165 and 75 per resolution element) for
July 6 and 7, respectively. The lower S/N ratio on the second night
spectrum reflects the poorer photometric conditions, and does not affect
the H$\alpha$ line analysis.

The data were reduced using the {\it Mauna Kea Echelle Extraction} -
MAKEE (T. Barlow) program for HIRES. It is optimized for the spectral
extraction of single, unresolved point sources. Data reduction was done
independently in a variety of ways in order to check for systematics in
the procedure.  The continuum was fit individually to each echelle order
(covering $\sim100$\AA) using a hot star observed during the same night
and 5th order polynomials.  The K3 giant spectra in each order were then
divided by the continuum fits to normalize the spectra.

We tested for and excluded such potentially confusing systematic effects
as faint nearby stars, sky background, and scattered light.
Contamination from the lens itself cannot be eliminated since they are
separated by $\sim 1\,$mas.  However, this contamination is expected to
be small since the source is both intrinsically bright and highly
magnified.

\section{Results}

\begin{figure}
\plotone{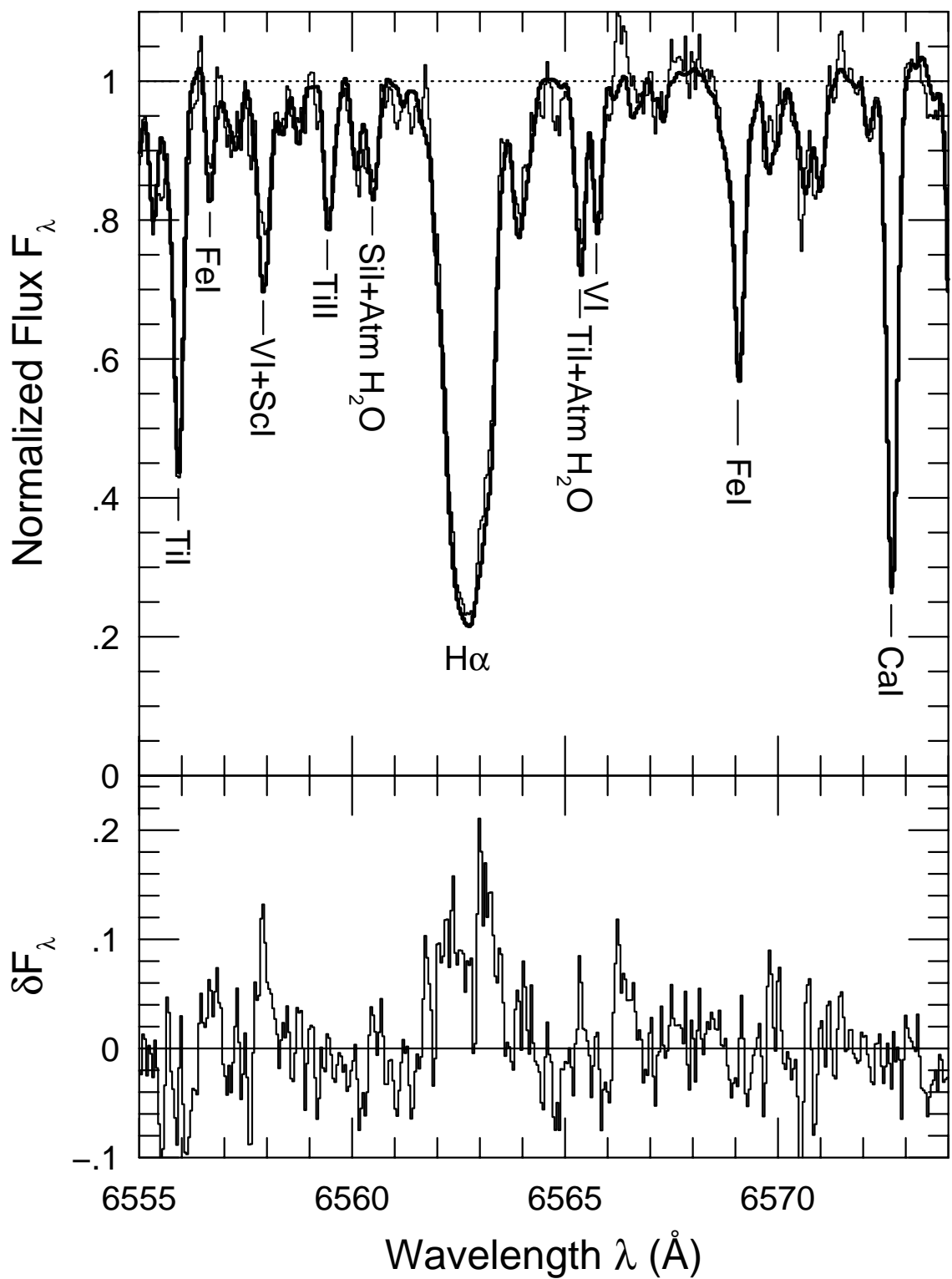}
\caption[junk]{\label{fig:two} Upper panel: HIRES spectra of EROS BLG-2000-5 
on UTC 2000 July 6 ({\it bold}) and 2000 July 7 ({\it solid}) around
H$\alpha$ (one order).  Lower Panel: Fractional difference in the
lines as a function of wavelength: $\delta
F_{\lambda} = 2(F_{\lambda,6} -
F_{\lambda,7})/(F_{\lambda,6}+F_{\lambda,7})$, where $F_{\lambda,6}$ and
$F_{\lambda,7}$.}
\end{figure}

Figure \ref{fig:two} shows the portion of the spectra centered on the
H$\alpha$ absorption line from the two nights.  The spectra were first
each normalized to a mean continuum of unity.  The spectra were aligned
in wavelength by removing a small residual shift of 0.256 pixels from
the July 7 spectrum, and the resolutions were matched by convolving the
July 6 spectrum with a 2.46 pixel FWHM Gaussian.  The wavelength scale
is 47m\AA/pixel.  The upper panel shows the spectra themselves, with key
lines identified, while the lower panel shows the fractional change in
line flux with wavelength, $\delta F_{\lambda} = 2(F_{\lambda,6} -
F_{\lambda,7})/(F_{\lambda,6}+F_{\lambda,7})$, where $F_{\lambda,6}$ and
$F_{\lambda,7}$ are the normalized fluxes (shown in the upper panel) for
July 6 and July 7 respectively.

\begin{figure}
\plotone{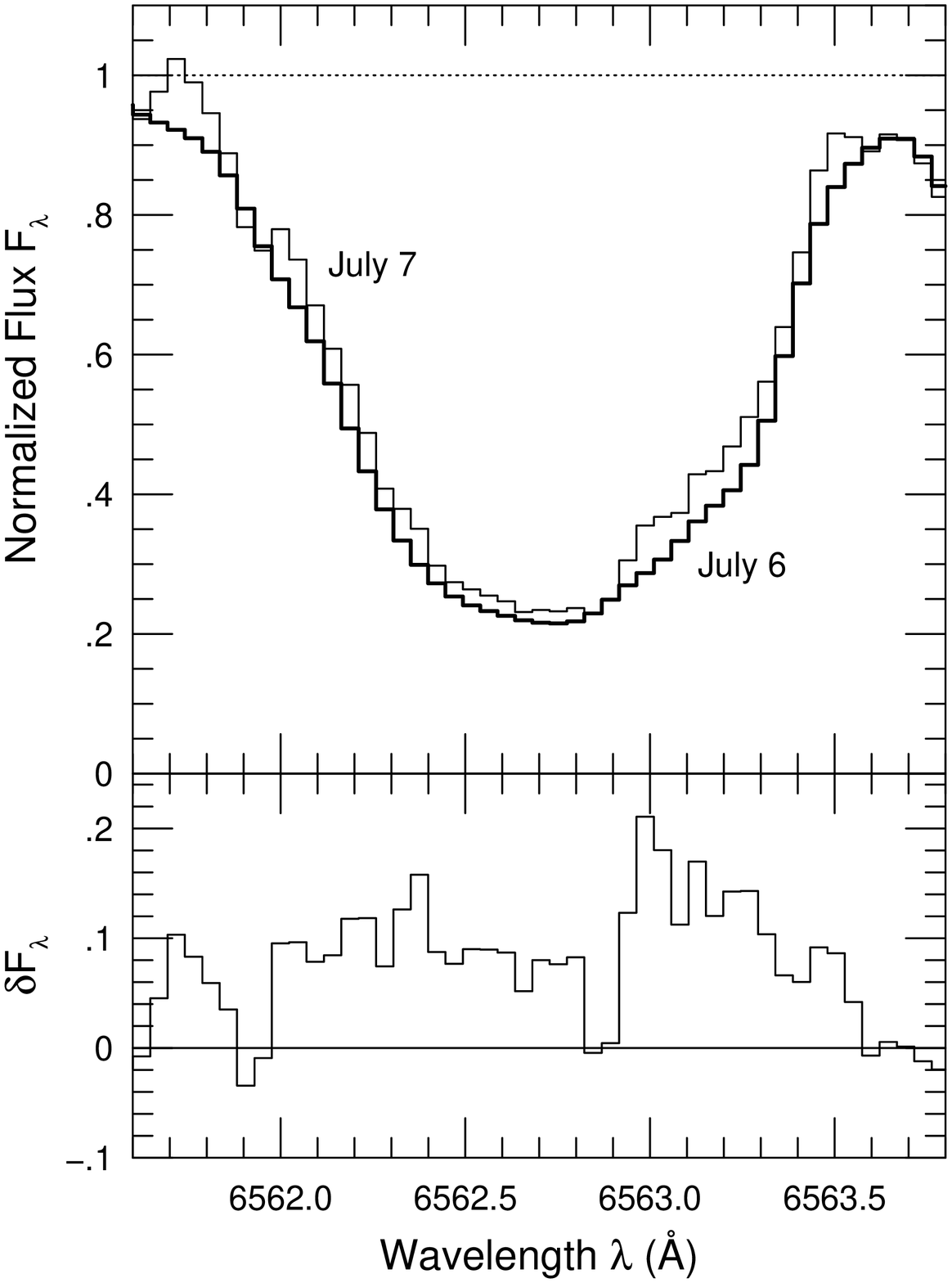}
\caption[junk]{\label{fig:three}
Detail of Figure \ref{fig:two} centered on the H$\alpha$ line.
}
\end{figure}

Figure \ref{fig:three} provides an expanded view of Figure \ref{fig:two}
for the H$\alpha$ line alone.  The fractional change in the line fluxes
as a function of wavelength is reasonably consistent with a constant
value, $\sim0.1$.  The equivalent widths of the lines are
$944\pm2$\,m\AA\ and $869\pm6$\,m\AA\, on July 6 and 7, respectively,
measured from the nominal continuum level in each spectrum.  These
equivalent widths differ by $8.3\%\pm 0.7\%$.

Since the two spectra were taken a day apart and the second-half
crossing time was predicted by PLANET to be about 2 days, the source had
changed its separation from the caustic by $\Delta \eta\sim 0.5$ in the
interval between the observations.  Hence, these spectra correspond
roughly to the magnification profiles labeled $\eta=0.25$ and
$\eta=0.75$ in Figure \ref{fig:one}.  That is, the July 7 spectrum is
concentrated near the limb, while the July 6 spectrum covers a much
broader range of stellar radii.  Thus, one expects that the July 7
spectrum reflects conditions higher in the atmosphere, implying a cooler
effective temperature, and so weaker hydrogen lines.  This is exactly
what is seen.  It also agrees with the predictions of Heyrovsk\'y et
al.\ (2000), although precise comparison is not possible, in part
because they considered point-lens caustics rather than binary-lens
caustics, and in part because they were modeling M stars rather than K
stars.

The behavior of adjacent spectral lines indicates that the difference we
see in the H$\alpha$ line is real and not an artifact of the reduction
process or other systematic effects.  In particular, the
\ion{Ca}{1}\,$\lambda$6572.9\AA\ line and the
\ion{Fe}{1}\,$\lambda$6569.2\AA\ line, which both lie in the same order as
H$\alpha$ (see Fig.\ref{fig:two}), show no detectable difference in
equivalent width between the two nights (228.4$\pm$2.3\,m\AA\ and
228.8$\pm$2.7\,m\AA\ for \ion{Ca}{1}, and 142.0$\pm$3.0\,m\AA\ and
149.3$\pm$3.3\,m\AA\ for \ion{Fe}{1}).  Not all lines are unchanged, for
example the \ion{V}{1}+\ion{Sc}{1} blend at 6558\AA\ (111.4$\pm$2.2\,m\AA\
compared to 77.8$\pm$4.7\,m\AA).  What is important is that while some
differences are seen, we do not see systematic differences among {\it all}
lines.  

\section{Conclusions}

We have presented spectra of a K3 giant taken on two different nights
during the second caustic crossing of the microlensing event EROS
BLG-2000-5.  The spectrum from the second night tends to be dominated by
light from the limb of the source, while the spectrum from the first
night covers a much wider range of radii.  Naively, one therefore
expects the second spectrum to reflect cooler conditions and so to
display weaker hydrogen lines.  The difference spectrum agrees with this
qualitative prediction.

To make more quantitative predictions for changes in this line (and in
all other spectral features covered by our spectra), two additional
inputs will be required: 1) the event caustic geometry, and 2) the
limb-darkening profile of the source.  Both can be obtained from
detailed modeling of the photometric light curve. It is clear from
Figure \ref{fig:one} that knowing the geometry, particularly $\eta_{\rm
July\ 6}$ and $\eta_{\rm July\ 7}$, is critically important.  The effect
of $A_3$ is to put a small uniform pedestal under all curves in Figure
\ref{fig:one} and is therefore not as crucial.  The effect of limb
darkening can be quite significant, as can be seen from the limb
darkening of a K2 giant shown in Figure 8 of Albrow et al.\ (1999).

When these various parameters are measured, we invite atmosphere
modelers to combine them with their atmosphere models to predict the
difference spectra.  It will then be possible to directly compare these
predictions with our data.

\begin{acknowledgements}

We thank the EROS, MPS, and PLANET collaborations for providing the
microlensing event alerts that made these observations possible.  We
thank Fred Chaffee, director of the W.M.\ Keck Observatory, for bringing
the alerts to the attention of the observers, and Don Terndrup for
useful suggestions.  Data were obtained at the W.M.\ Keck Observatory,
which is operated as a scientific partnership among the California
Institute of Technology, the University of California, and the National
Aeronautics and Space Administration.  The Observatory was made possible
by the generous financial support of the W.M.\ Keck Foundation.  This
work was supported by NSF grants AST-9727520 and AST-9530619, NASA grant
NAG5-7589, and a grant from Le Minist{\`e}re de L'{\'E}ducation
Nationale de la Recherche et de la Technologie.

\end{acknowledgements}

\subsection*{Note Added in Proof:}

After we received the referee's comments to the original version of this
paper, we learned of the work of Albrow et al.\ (astro-ph/0011380)
reporting complementary low-resolution observations of the same caustic
crossing.

\newpage

\newpage

\begin{references}

\reference{} Afonso, C., et al.\ 2000, \apj, 532, 340
\reference{} Albrow, M.D., et al.\ 1999, \apj. 522, 1011
\reference{} Albrow, M.D., et al.\ 2000, \apj, 534, 894
\reference{} Albrow, M.D., et al.\ 2001, ApJ, 549, 000 (astro-ph/0004243)
\reference{} Alcock, C.\ et al.\ 1997, \apj, 491, 436
\reference{} Alcock, C.\ et al.\ 2000, \apj, 541, 270
\reference{} Gaudi, B.S., \& Gould, A.\ 1999, 513, 619
\reference{} Gould, A.\ 1997, 483, 98
\reference{} Heyrovsk\'y, D., Sasselov, D.D., \& Loeb, A.\ 2000, \apj, 
submitted (astro-ph/9902273)
\reference{} Ignace, R., \& Hendry, M.A.\ 1999, \aap, 341, 201
\reference{} Lennon, D.J., Mao, S., Fuhrmann, K., \& Gehren, T.  1996, 
             \apj, 471, L23
\reference{} Maoz, D., \& Gould, A.\ 1994, \apj, 425, L67
\reference{} Popper, D.M., Andersen, J., Clausen, J.V., \& Nordstrom, B.\ 1985,
             \aj, 90, 1324
\reference{} Uitenbroek, H., Dupree, A.K., \& Gillilan, R.L.\ 1998, \aj,
             116, 2501
\reference{} Valls-Gabaud, D.\ 1996, in IAU Symp.\ 173, Astrophysical
Applications of Gravitational Lensing, ed. C.S.\ Kochanek \& J.N. Hewitt
(Dordrecht: Kluwer), 237
\reference{} Valls-Gabaud, D.\ 1998, \mnras, 294, 747
\reference{} Vogt, S., et al. 1994, SPIE, 2198, 362
\reference{} Witt, H.J.\ 1995, \apj, 449, 42

\end{references}
\end{document}